\documentclass[structabstract]{aa}
\usepackage{graphicx}
\usepackage{txfonts}
\usepackage{latexsym}
\usepackage{bm}

\begin{document}

\title{\textit{z=1} Multifractality of Swift short GRBs?}
\author{Fabrizio Tamburini}

\offprints{F. Tamburini, \email{fabrizio.tamburini@unipd.it}} 

\titlerunning{Fractal - multifractal distribution - Angular distribution of GRBs}
\authorrunning{F. Tamburini}
\institute{Universit\`a di Padova, Dipartimento di Astronomia,
Vicolo dell'Osservatorio 3, I-35122 Padova, Italy
}

\date{Received ; accepted }

\abstract
{}
{We analyze and characterize the angular distribution of selected samples of gamma ray bursts (GRBs) from Batse and Swift data to confirm that the division in two classes of short- and long-duration GRBs correspond also to the existence of two distinct spatial populations.} 
{The angular distribution is analyzed by using multifractal analysis and characterized by a multifractal spectrum of dimensions. Different spectra of dimensions indicate different angular distributions.}
{The spectra of dimensions of short and long bursts indicate that the two populations have two different angular distributions. Both Swift and BATSE long bursts appear to be homogeneously distributed in the sky with a monofractal distribution. Short GRBs follow instead a multifractal distribution for both the two samples. Even if BATSE data may not give a secure interpretation of their angular distribution because of the instrumental selection effects that mainly favor the detection of near GRBs, the results from Swift short GRBs confirm this behavior, also when are included GRBs corrected by the redshift factor.
The distributions traced by short GRBs, up to $z=1$, depict a universe with a structure similar to that of a disordered porous material with uniformly distributed heterogeneous irregular structures, appearing more clustered than what expected.}
{}


\maketitle

\section{Introduction}

Gamma Ray Bursts (GRB) are catastrophic explosions of cosmological origin that illuminate the sky once or twice a day with relatively short, intense, flashes of $\gamma$-rays on the order of the MeV and duration that ranges from $10^{-3}$ to about $10^3$ seconds.
Until the launch of Swift, the most widely accepted taxonomy of GRBs is the division between short-hard and long-soft bursts, after Dezalay (1992) and Kouveliotou (1993), who found in BATSE data a bimodal distribution in the burst duration defined as the time it takes the $90\%$ of the flux to arrive ($T_{90}$), with respect to the local time of the detector. These two distinct GRB distributions are separated by a minimum located at $T_{90} \simeq 2$ s. This sharp division is supposed to be also due to a selection effect of the instrumentation onboard the satellite. 
In fact, new Swift observations permit to include in the classification scheme, as short bursts, also some distant events having apparently $T_{90}>2$ s with respect to Swift's proper time. The determination of the redshift permits in fact to correct the $T_{90}$ by the relativistic effects (\cite{che97,ruf09}). 
This suggests the need of a more solid classification of GRBs on the basis on a broader set of criteria, beyond the mere burst duration and the verification of the present taxonomy of long/short GRBs (\cite{don06,zha07,blo08,bel08}).

Long and short GRBs are thought to have been generated by different progenitors. Anyway, different progenitors are thought to produce a black hole accreting material from a disc or from a torus with the emission of gamma rays (\cite{ghi09}). The basic model for long-duration GRBs, related to the catastrophic release of energy from the collapse of massive stars (\cite{woo01,fry01}), has received a strong support from the observations of their X (\cite{geh08,nys08}), optical and radio counterparts, and the association with supernova detections, of which the GRB980425/SN1998bw association was the first clear example (\cite{gal98,kul98,vpa99}). 
Thus, X-ray flashes, associated to long-duration GRBs, are probably produced by the highly-relativistic jets ejected in core-collapse supernova explosions. 
The relativistic fireball model (\cite{goo86,pac86,pac90,cas01}) accounts for a reasonable description of the observed afterglow spectrum, powered by the synchrotron emission of electrons accelerated in a relativistic shock with an estimated total energy budget roughly the same order of magnitude as that of supernovae Ib/c. 
The discovery of the slowly fading X-ray, from optical and radio afterglows of GRBs and the identification of host galaxies at cosmological distances, gave further support to describe their progenitors as the product of the evolution of short-lived massive stars in different cosmological epochs, whose detection is limited by the BATSE threshold that sets the observational limit at distances slightly larger than $z \sim 4$ (\cite{wij97,bla00}). Now the new data from Swift passed the barrier of $z=8$ (\cite{sal09}).

Short GRBs are thought to be produced by highly - relativistic jets ejected in different processes, as  during neutron star - neutron star (NS-NS) or black hole - neutron star (BH-NS) binary mergers  (\cite{nar92}), whose averaged redshift distribution, in the Swift-era, seems to be $\langle z \rangle \sim1.0$ (see e.g. Magliocchetti, Ghirlanda \& Celotti, 2003, Tanvir et al. 2005, and Ghirlanda et al. 2006).
Anyway, recent  Swift and HETE-2 observations put in evidence the presence of clear similarities with the afterglows detected in the correspondence of long GRBs, thanks to the detection of X-ray and optical afterglows. Other similarities found  between the initial stage of the spectra of short and long GRBs together with the presence, in some cases, of X-ray flares, seem to indicate the presence of a common mechanism during the first few seconds (\cite{bar05,cow07}). 
Short GRBs have been observed mostly in elliptical galaxies, but even less frequently also in nearby irregular and in star-forming galaxies, confirming as progenitor the binary merging scenario (\cite{mes06, nar01,bel08}) also through the mechanism of \textit{star swapping} (\cite{gri06}). A similar behavior is expected for NS-BH binaries (\cite{pac91}).
The angular and spatial distributions of BATSE GRBs look quite isotropic (\cite{bri96}), with only very small  anisotropies observed in the angular distribution (\cite{mes99}); short- and long- GRBs in BATSE catalog actually show two different angular distributions (\cite{vav08,bal09}), but the connection to the hypothesis of the instrumental selection effect is not still completely clear.

In this paper we characterize the angular distributions of the two classes of GRBs with multifractal analysis from a selected sample of BATSE and Swift observations, taking also in consideration the redshift relativistic effects and  compare the results obtained from the two catalogues.
In Chapter 2 are described the mathematical basis of the method. 
In Chapter 3 are estimated the fractal/multifractal dimensions by determining the distribution moments of the multifractal spectrum from the second up to the tenth order and then draw the conclusions.


\section{Fractal / Multifractal analysis of the angular distribution}

Being GRBs associated to galaxies, they should trace the angular distribution of their host galaxies at  distances slightly larger than those estimated with classical Supernovae, i.e. distances the order the Gpc (\cite{pac86,uso75}), across which inhomogeneity in the distribution of luminous matter should be averaged out from the Mpc scale.

Mandelbrot and Peebles (1980) first used fractal geometry to describe the angular and spatial distribution of galaxies. In their work they used L\'evy-Rayleigh random paths, i.e. infinite-variance, stable, generalized random walks in which the step lengths are described by a tailed probability distribution. Galaxies are placed at the steps of a random walk and each galaxy is randomly connected with another in the close proximity, like mimicking the random motion of a fly in the air passing through each given point. 
The power-law constraint of the motion gives the fractal exponent of the random path and rules the distribution of the jump lengths and the direction of each jump is taken isotropically at random (\cite{mar02}).

Galaxy surveys show that the distribution of the luminous matter in the universe is more complicated than that of a single fractal. The distribution of galaxies has instead multifractal properties (\cite{pie87,cel01}), with a tendency to cluster following the well known peculiar structures made by patterns of voids and filaments (\cite{kur99}).
The CfA survey of nearby galaxies, for example, presents scaling properties with a correlation dimension in the interval $D_{2}\simeq 1.3-1.4$. The two-point correlation function and the power spectrum analysis indicate that the distribution of galaxies at very large scales becomes homogeneous and isotropic like the X-ray background emitted by active galactic nuclei (AGNs), in agreement with the Cosmological Principle (\cite{pee93}). 

Multifractal scaling analysis revealed the presence of aggregated structures also in some samples of galaxies with distances larger than $r\sim 20$ Mpc. At scales larger than $30$ Mpc, the angular distribution of luminous matter tends to homogeneity. According to \cite{pie87} the distribution of luminous matter has multifractal properties with fractal dimensions varying from $d=1.23$, for the nearby galaxies, up to the value $d\simeq 2$ at very high redshifts.
Usually, multifractal distributions are present when a structure has different fractal dimensions on different parts of the support; in other words, when spatial correlations are present and change the geometrical shape of the distribution at different scales (\cite{fal90}).  Those distributions cannot be adequately described by a geometrical support with a single fractal dimension, but require instead a whole spectrum of dimensions.

To characterize the angular distribution of GRBs in terms of multifractals, we use the ``\textit{method of moments}'' that estimates the fractality of the distribution by calculating the multifractal spectrum of generalized dimensions $D_q$ in a given range (\cite{fal90,fed88}). 
Consider $N$ GRBs in a box of size $L$ divided into cells of size $r$. The sample is said to have a non-null $q-th$ moment $\xi_q$ on the scale $r$ \textit{iff}
\begin{equation}
\xi_q(r)=\sum^{L/r}_{i=1}\left( \frac{n_i}{N}\right)^q \neq 0
\,\,\,\,\,\, \mathrm{and} \,\,\,\,\,\, (q>2)
\end{equation}
where $n_i$ is the number of objects present in the i-th cell. Higher moments reflect the emergence of structures present in the denser regions. From the scaling relation conducted on a two-dimensional section we obtain (Kurokawa et al. 2001),
\begin{equation}
\xi_q(r)\sim r^{(q-1)D_q} \,\,\,\,\,\, (r_{min}<r<r_{max}).
\end{equation}
where the coefficients $D_q$ belong to the spectrum of generalized dimensions.
$D_0$ is the {\em capacity dimension}, $D_1$ is the {\em
information dimension}, and $D_2$ is the {\em correlation dimension}.

To estimate the multifractality of the distribution one has to determine each generalized dimension $D-q$ as a function of the $q$-th moment. One gives evidence to the dependence of each of the fractal dimensions from the moments by drawing the $q-D_q$ plot, which is based on the Lipshitz-H\"older exponent (\cite{ben98, gol97}). In the $q-D_q$ plot one defines \textbf{monofractals} those simple fractal structures described by only one dimension, $D_f$, and in the plot all the $D_q$'s result equal to $D_f$.
In the general case of a \textbf{multifractal}, instead, $D_q$ is usually decreasing for higher and higher values of the moments $q$: if $q>q'$, then $D_q \leq D_{q'}$ until converging to the asymptotic value $D_\infty$ for a distribution of infinite objects. 

To characterise with a good precision the multifractal distribution of a sample of object in space one usually does not need to calculate the generalized dimensions up to the limit dictated by the number of points in the space. For our purposes a good estimation was obtained by taking as upper limit $q=10$. 

\section{Results and discussion}

From Swift recent observations we selected a sample of $444$ GRBs without ambiguity of classification, $52$ of which are classified as short- GRBs after the correction of $T_{90}$ from the relativistic effects (data updated at 2009-09-15).
From BATSE 4 catalogue, instead, the total number of GRBs used in our analysis is $1843$, $1447$ of which are classified as long-GRBs. 

To verify whether the classification in the two subgroups actually correspond to two distinct populations in both the sets of data collected by Batse and Swift, we also performed a series of tests of the angular distributions in different sub-samples, by randomly choosing from our data a sets of either short or long GRBs and mixing them together, to see whether GRBs might be discriminated without choosing ``\textit{a priori}'' the two classes following the burst duration time distribution. 
The two classes of long/short GRBs emerged anyway from the fractal analysis of their angular distribution in the sky. More precisely, by progressively mixing the population of short GRBs with a growing sample of randomly chosen long GRBs, the multifractal dimension converges to that of a homogeneously distributed monofractal with dimension $d=2$. This simple test corroborates the actual sub-division of GRBs in the two short/long populations for both the Batse and Swift data and also when the two datasets are mixed together.

\begin{figure}
\includegraphics[angle=0,width=8.5cm, keepaspectratio]{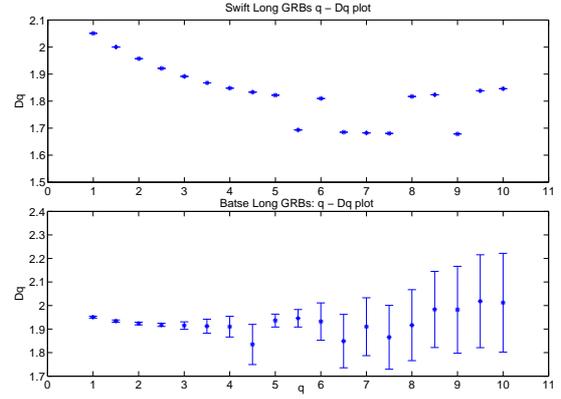}
\caption{\textbf{Upper panel}: $q-D_q$ plot of Swift \textit{long} GRBs. The long bursts follow an angular homogeneous distribution. The moments $D_q$ tend to decrease slowly from $D_1\sim2$ to $D_q\sim1.8$ with a rise of the fractal dimension after $D_q=8$. The spectrum of dimensions is that of a two dimensional brownian motion with fractal dimension $2$.
\textbf{Lower panel:} also BATSE 4 long bursts present a very similar homogeneous structure, with a mean fractal dimension $D \simeq 1.98$. In this case, a deviation in the distribution due to the moments $4<D_q<6$,  is present.} 
\label{long}
\end{figure}

We now present and discuss the results obtained for each different class of GRBs:

\textbf{Long bursts}:  Fig. (\ref{long}) clearly shows the uniform angular distribution of both the samples of Swift and BATSE 4 selected data.
The sample of data from Swift shows a homogeneous distribution. In this case, the moments $D_q$ tend to decrease slowly from $D_1\sim2$ to $D_q\sim1.8$ and the fractal dimension starts increasing after the moment $q=8$. In the plot are also present some sporadic and discontinuous jumps down to $D_q\sim 1.7$ that deviate from the main smoothness, a behavior that might be due to numerical errors in the determination of certain $D_q$'s . In any case, this does not give rise to misunderstandings to the interpretation of the global behavior observed in the plot. What is seen is a typical example of a stochastic homogeneous distribution, similar to that of a fractional brownian motion (FBM) having dimension $D=2$, as expected from distant sources that homogeneously distribute according to the cosmological principle. 
The errorbars of the $D_q$s' are calculated from the error propagation of the statistical uncertainty of the position of each GRB and from the instrumental errors. In the case of Swift data, errorbars are too small to be visible in the graph. 

By analyzing BATSE data, we find a different behavior in the interval $4<D_q<6$, where is seen a small bump. Also in this case, within the experimental errors, the angular distribution is characterized by an almost constant fractal dimension fluctuating within the interval $D_q= [1.8- 2]$ that approximately describes the angular distribution of a uniform structure. 
Having found a similar result in both the datasets, being the redshift scale around the Gyr (as confirmed by the redshift values obtained for most of the bursts), long GRBs distribute according to the the Cosmological Principle, in which the Universe is completely homogeneous and isotropic at large scales. 

\begin{figure}
\includegraphics[angle=0, width=8.5cm, keepaspectratio]{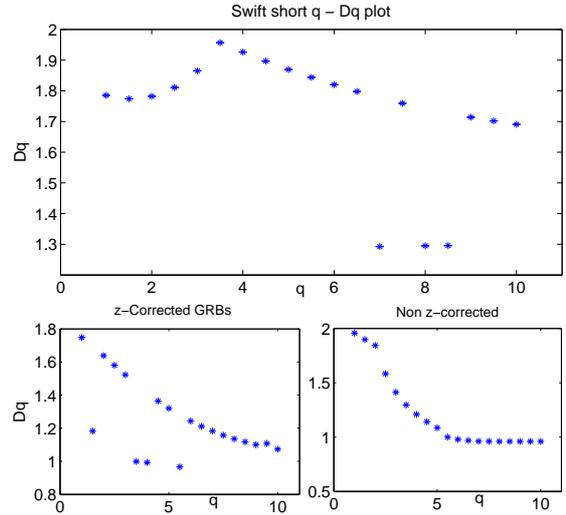}
\caption{$q-D_q$ plot of Swift \textit{short} GRBs. \textbf{Upper panel}: spectrum of fractal dimensions of the angular distribution of the whole sample. This behavior with a maximum around $D_q=3$ clearly indicates a multifractal homogeneous distribution very close to that of a multifractional brownian motion. \textbf{Lower left panel:} dimension moments of the z-corrected GRBs with relatively high redshift. The structure is clearly multifractal.  \textbf{Lower panel:} also those GRBs with already $T_{90}<2$ sec. show a multifractal distribution, but with an asymptotic convergence to $D_q=1$. In this case the structure traced by the GRBs appears more clustered.} 
\label{shortsw}
\end{figure}

\textbf{Short bursts} actually represent a more articulated and complicated set of phenomena with still unknown aspects. 
Fig. \ref{shortsw} shows that the whole sample of $53$ Swift short bursts (together with the apparently-long bursts then corrected by $z$ and those with $T_{90}$ already below $2$ s) is characterised by a spectrum of dimensions that decreases for increasing $q$'s, which is the signature of a multifractal distribution. 

Short GRBs look angularly distributed as they were driven by a multifractional brownian motion (mBM) of mean index $H=0.68$, a distribution that starts with a clustered shape then evolving towards homogeneity at larger scales. The mBM is an extension of the fractional Brownian motion in the sense that the path regularity can vary with time as is seen in the evolution of anomalous diffusion processes. The Universe traced by short short GRBs exhibits a multifractal structure, correlated at lower scales and evolving to homogeneity at larger distances. This structure appears to be similar to a disordered porous material, that presents heterogeneous structures, or even irregular in a uniform sense. This type of foam contains multiple, nested natural length scales or continuously evolving scales, while moving to higher redshifts (\cite{lim02}).

The further division in two sub-classes with respect to the $T_{90}$-corrected by the redshift and those that were already below the $2$ seconds, show that the two sub-samples follow two mutually different multifractal distributions and the sub-class of the z-corrected short GRBs presents a higher fractal dimension $q_0$ which is close to that of a homogeneous distribution, confirming the behavior expected from a mBM (see the two lower panels in Fig. \ref{shortsw}). 

\begin{figure}
\includegraphics[angle=0, width=8.5cm, keepaspectratio]{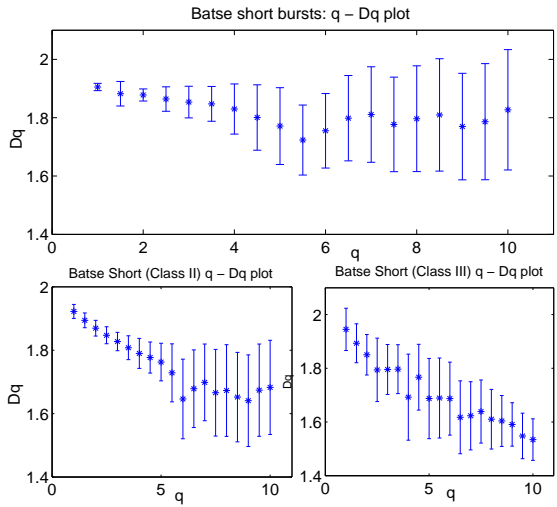}
\caption{\textbf{Upper panel:} $q-D_q$ plot of BATSE 4 \textit{short} GRBs (Class I  + Class II) 
present a fractal structure with dimension fluctuating around the value $\gtrsim 1.8$.
 \textbf{Lower left panel:} $q-D_q$ plots of BATSE 4 short bursts Class II and
Class III alone clearly follow a multifractal distribution.} 
\label{short}
\end{figure}

The whole sample of BATSE short GRBs shows an articulated spectrum of the general dimension $D_q$, with a smoother behavior with respect to the whole sample of BATSE long GRBs, as reported in the upper panel of Fig. \ref{short}. The capacity dimension $D_0$ and the correlation dimension $D_2$ present similar values close to $D_q\simeq1.9$ that are smaller than those of long GRBs. This demonstrates that short GRBs have a different multifractal and more clustered distribution than the class of long ones. This difference is better put in evidence if BATSE GRBs are divided in the two subclasses indicated by \cite{muk98}: 
Class II with short/faint/hard bursts and Class III with intermediate/intermediate/soft bursts (we remind that Class I correspond long GRBs), as drawn in the lower panels of Fig \ref{short}.
Here, the generalized dimension $D_q$ of both II and III classes alone indicates that the angular distribution is clearly multifractal, similar to that seen in the galaxy distribution at low redshifts. 

Concluding, even if this effect could be ascribed to a selection effect in BATSE 4 data, anyway Swift results confirm the existence of two different angular distributions associated to the two classes of short and long GRBs. 
In both Swift and BATSE data sets long bursts are homogeneously angularly distributed. Short GRBs, instead, trace a distribution which appears different to that expected from the clustering of luminous matter in the Universe around the redshift value $z\sim 1$. We in fact still clearly see a multifractal distribution in a structure that should be already homogeneously-distributed.
By taking in account the fact that the candidate progenitors of short GRBs are thought to migrate away from their initial positions, we should also expect an additional blurring effect on the detected structure that should smear out the distribution towards the total homogeneity, which has not been observed. This suggests that either GRBs are not good tracers for the matter distribution or that the Universe traced by GRBs might appear more clustered at redshifts $z \lesssim 1$ than expected.

\section{Acknowledgments}
The author would like to acknowledge Massimo della Valle for the helpful discussions and suggestions.

\end{document}